\documentstyle[sprocl]{article}
\input{psfig}
\def\Journal#1#2#3#4{{#1} {\bf #2}, #3 (#4)}
\def\PLB{{\em Phys. Lett.}  B}
\def\PRL{\em Phys. Rev. Lett.}
\def\PRD{{\em Phys. Rev.} D}

\def\lsim{\mathrel{\rlap{\lower4pt\hbox{\hskip1pt$\sim$}}
    \raise1pt\hbox{$<$}}}         %less than or approx. symbol
\def\gsim{\mathrel{\rlap{\lower4pt\hbox{\hskip1pt$\sim$}}
    \raise1pt\hbox{$>$}}}         %greater than or approx. symbol

\def\be{\begin{equation}}
\def\ee{\end{equation}}
\def\bea{\begin{eqnarray}}
\def\eea{\end{eqnarray}}

\begin{document}
\title{$^3$He Transport and the Question of Nonstandard Solar Models}
\author{W.C. Haxton}
\address{Institute for Nuclear Theory, Box 351550, and Dept. of Physics,
Box 351560 \\
University of Washington, Seattle, WA 98195-1550}

\maketitle\abstracts{
I consider phenomenological changes in the standard solar model with the goal 
of testing recent claims that the solar neutrino puzzle requires new particle 
physics.  The assumption of a steady-state sun producing the correct 
luminosity and governed by standard microphysics appears to leave only one 
nonstandard solar model possibility open, slow mixing of the solar core on 
timescales characteristic of $^3$He equilibration.  The conjecture of 
such mixing 
raises striking physics issues connected with the 
standard model $^3$He instability and the possibility that the $^3$He 
abundance gradient might allow 
the sun's early convective core to persist.  While helioseismology might
eventually rule out such a ``model", contrary to one recent claim I will 
argue that the helioseismology of a mixed two-fluid sun is as yet far from clear.  
Finally, I conclude by stressing that $^3$He-driven slow mixing is not being 
proposed as a solution of the solar neutrino problem, but as an example of a 
possibility that has not been quantitatively modeled and yet could produce 
neutrino fluxes far closer to experiment than the standard solar model.  Thus, 
despite the attractiveness of neutrino oscillation solutions, 
astrophysical explanations of the solar neutrino problem are not yet, in my 
view, ruled out definitively.}

In this talk I would like to summarize some recent work, done in collaboration 
with Andrew Cumming, on the possibility of an astrophysical solution to the 
solar neutrino problem [1].  It is widely appreciated that the results of the 
$^{37}$Cl, SAGE/GALLEX, and
Kamioka II/III experiments are consistent with an unexpected
pattern of neutrino fluxes,
\begin{eqnarray}
\phi({\mathrm {pp}})  &\sim& \phi^{SSM}({\mathrm {pp}}) \nonumber \\
\phi (^7{\mathrm {Be}}) &\sim& 0 \\
\phi(^8{\mathrm B}) &\sim& 0.4 \phi^{SSM} (^8{\mathrm B})
\nonumber
\end{eqnarray}
where $\phi^{SSM}$ denotes the standard solar model [2] (SSM) value.
As $\phi(^8$B) $\sim T_c^{22}$ [3], where $T_c$ is the solar core
temperature,
the required reduction in this flux can be achieved by lowering $T_c$
to about 0.96 of the SSM value.
However, as $\phi(^7$Be)/$\phi(^8$B) $\sim T_c^{-10}$,
this flux ratio then increases, contradicting Eq. (1).
The difficulty of
simultaneously reducing $\phi(^8$B) and $\phi(^7$Be)/$\phi(^8$B)
has been established for broad classes of solar models [3-5],
leading many to favor nonastrophysical solutions
to the solar neutrino problem.

It is clear that no solar model
will give a perfect fit to the results of existing
experiments: the measurements are inconsistent with any
combination of undistorted $^8$B, $^7$Be, and pp neutrino fluxes at
a confidence level of about 2$\sigma$ [6].  Yet
a compelling argument for a resolution in terms
of new particle physics must rest on the more dramatic $\sim 5 \sigma$
discrepancy, illustrated in Fig. 1, that exists between experiment
and the
flux predictions of standard and nonstandard models.  Thus it
is important to determine whether a nonstandard model
might exist in which the naive $T_c$ dependence described above
is circumvented. 
\begin{figure}
\begin{center}
~\psfig{figure=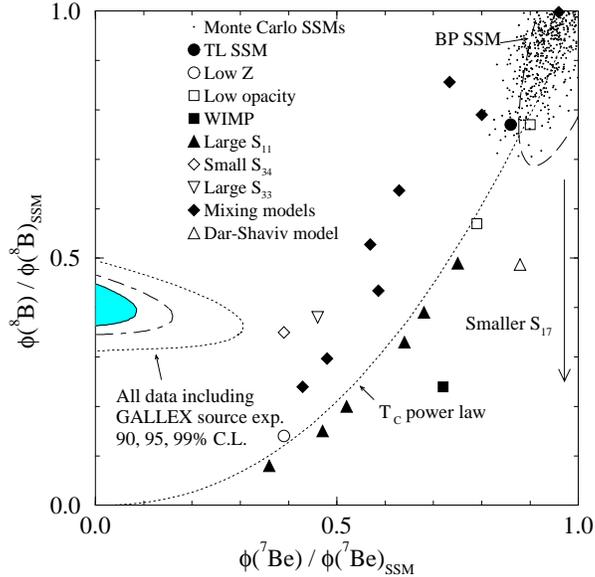,height=3in}~ 
\end{center}
\caption{A comparison of SSM 
(ellipse in 
upper right hand corner) and various nonstandard model neutrino flux 
predictions to experiment.  Notice that the theoretical results cluster along 
a trajectory corresponding to the naive $T_c$ dependence discussed in the 
text. (From Ref. [5])}
\end{figure}
The philosophy behind the calculations presented in Ref. [1] stemmed from a 
worry that, if a viable nonstandard model exists which is approximately 
compatible with 
the results in Eq. (1), its underlying physics might be subtle and thus 
difficult to anticipate.  This seemed to argue for a simple minded 
approach $\--$ changing the SSM phenomenologically $\--$ putting aside for 
the moment the 
more difficult issue of the underlying physical mechanism, in the hope that 
Eq. (1) might then lead us to the proper solution.  The procedures we followed 
are discussed in Ref. [1] and will not be repeated here.  But the basic 
approach was to search for solutions more consistent with Eq. (1) constrained 
by three conditions.  First, we retained all of standard nuclear and atomic 
microphysics, e.g., nuclear cross sections and opacities, because we felt 
current SSM ``best values" for these parameters were sensibly chosen, given 
the existing body of measurements.  Second, we required, to the extent 
possible in our phenomenological approach, that the modified ``models" 
reproduce the known solar luminosity.  Finally, we required the ``models" 
to be steady state, thus demanding where appropriate an equilibrium in the 
production and consumption of pp-chain ``catalysts" like D, $^3$He, and 
$^7$Be.  Note that the SSM requires such equilibrium locally, while we allowed 
more freedom in the ``model" by only enforcing this condition in the integral 
core abundances.

In the SSM the core $^3$He abundance profile is very distinctive, rising 
steeply as a function of the distance r from the center of the sun: the 
equilibrium abundance $X_3$ varies as T$^{-6}$, where T is the local temperature. 
 But the characteristic result of our phenomenological explorations was a 
somewhat cooler sun with a remarkably different $^3$He profile, one elevated 
by an order-of-magnitude, relative to the equilibrium value, at small r and 
depleted at large r.  Such an altered profile is compared to the SSM result in 
Fig. 2.   
\begin{figure} 
\begin{center}
~\psfig{figure=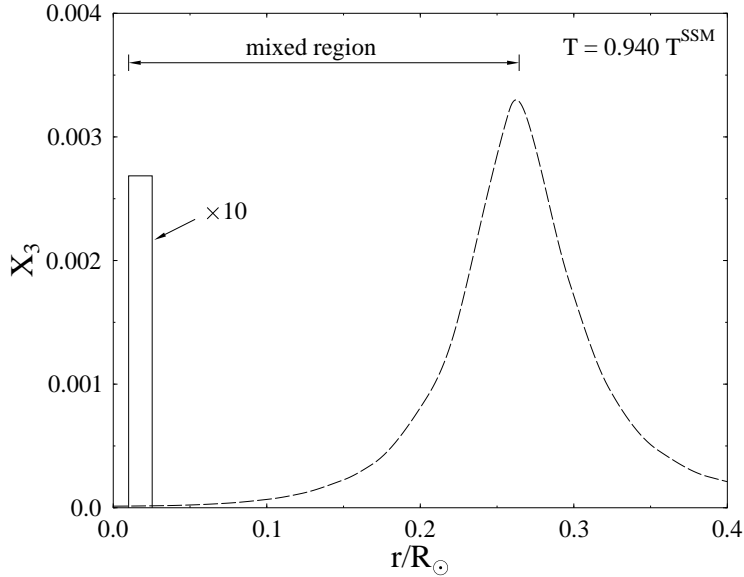,height=3in}~
\end{center}
\caption{The dashed line gives the 
SSM equilibrium $^3$He mass fraction rescaled for a cool sun (T = 0.94 
T$^{SSM}$).  The solid line is a modified $^3$He profile producing an 
equivalent $^3$He burning rate, the correct luminosity, and neutrino fluxes 
similar to those of Eq. (1).  (See Ref. [1]).}
\end{figure}
It is readily seen why such a change moves the neutrino flux predictions 
towards the results of Eq. (1).  First, a large fraction of the produced 
$^3$He is burned out of equilibrium at small r.  The ppI terminations are 
governed by the reaction $^3$He+$^3$He, which is quadratic in the $^3$He abundance, while 
the competing reaction $^3$He+$^4$He leading to higher energy neutrinos is linear.  Thus the rate of ppII+ppIII 
terminations relative to ppI terminations is reduced in direct proportion to 
the $^3$He excess, suppressing both the $^7$Be and $^8$B neutrino fluxes.  
However, when the reaction $^3$He+$^4$He does occur, short-lived $^7$Be is produced at 
small r, where the ambient temperature is high.  This favors ppIII 
terminations over ppII terminations, leading to a suppressed 
$\phi(^7$Be)/$\phi(^8$B) flux ratio.  The combined effects of the reduced 
(ppII+ppIII)/ppI and enhanced ppIII/ppII branching ratios yield a somewhat 
reduced $^8$B neutrino flux and a significantly reduced $^7$Be flux.

Now how should Fig. 2 be interpreted?  Clearly the nonstandard $^3$He profile 
represents something quite different from the static SSM profile with which 
it is compared in the figure: such an 
out-of-equilibrium profile can only result from core mixing of $^3$He on a 
timescale characteristic of $^3$He equilibration ($\sim$ 10$^7$ years in 
the outer core).  One soon concludes that the nonstandard profile depicts 
where $^3$He is burned in a postulated ``model", which may be quite different 
from the abundance distribution itself.  Such localized burning of $^3$He at 
small r can arise from a rather specific pattern of core mixing.
First, there must be a relatively rapid downward flow of
$^3$He-rich material from large r; the speed must be
sufficient to take a mass element well past the
usual equilibrium point, into a region where the rapidly
decreasing local lifetime of $^3$He finally results in sudden $^3$He
ignition.  This mass element, now depleted in $^3$He and
buoyant because of the energy release, must return to
large r sufficiently slowly to allow the p+p reaction to
replenish the $^3$He.  This flow is depicted in Fig. 3.  As
we are assuming a steady-state process in which any mass
element is roughly equivalent to any other,
each mass element must, on average, remain within
a radial shell bounded by r and r+dr for a time proportional
to the mass dM(r) contained within that shell.  This condition
would be satisfied if the slow upward flow is broad
with a local velocity inversely proportional to dM(r) -
the kind of flow that would result from displacement from below.
Such upward flow will produce a positive $^3$He gradient,
as in the SSM; but the upward flow must be sufficiently fast to
keep the $^3$He below its local equilibrium value to prevent
burning at large r.
To keep the circulation steady, the rapid downward flow clearly
must be localized, e.g., perhaps in narrow plumes.  This flow was simulated 
numerically in Ref. [1].  The derived $^3$He burning pattern and neutrino 
fluxes ($\phi(^8$B) $\sim$ 0.4 $\phi^{SSM} (^8$B), $\phi(^7$Be)/$\phi(^8$B) 
$\lsim \phi^{SSM}(^7$Be)/$\phi^{SSM} (^8$B)) emerged for a variety of downward 
$(\tau_\downarrow \sim$ few $\cdot 10^6$ y) and upward $(\tau_\uparrow \sim$ 
few $\cdot 10^7$ y) flow time scales, provided the mixing encompasses most of 
the energy-producing core. 
\begin{figure}
\begin{center}
~\psfig{figure=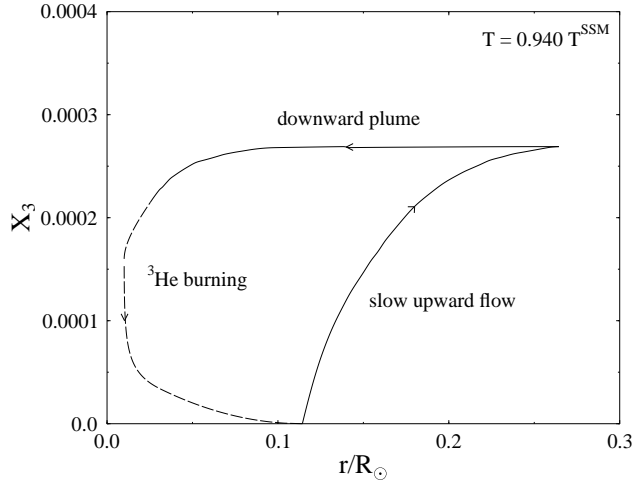,height=2.5in}~
\end{center}
\caption{A schematic $^3$He 
circulation pattern suggested by Fig. 2.  The solid line represents 
descending, localized $^3$He-rich plumes (downward arrow) and broad, slow 
restoring flow (upward arrow).  The dashed line, representing the process of 
$^3$He ignition, buoyancy, and subsequent cooling, has not been modeled 
numerically.}
\end{figure}
It is notable that a simple flow pattern like that of Fig. 3 can produce the 
$^3$He burning pattern of Fig. 2:  the latter was deduced phenomenologically, 
so it is not obvious that it will necessarily arise from any mixing 
pattern.  Furthermore, there exist some speculative but rather intriguing 
corrections to well known idiosyncracies of the SSM:

1)\  In their work on the ``solar spoon", Dilke and Gough [7] showed that the 
SSM $^3$He gradient implied an instability to large amplitude perturbations:  
the energy released by enhanced $^3$He burning can exceed the work against 
gravity required to push a volume element at large r through the denser 
material below.  They speculated that 
this overstability could trigger periodic mixing of the core.  Our 
phenomenology suggests an interesting variation on the solar spoon mixing 
mechanism.  In the case of 
the continuous flow depicted in Fig. 3, the core would remain homogeneous in H 
and $^4$He while still permitting a $^3$He gradient, an amusing change in  
the solar spoon as the flow would be nearly adiabatic.  Large-scale adiabatic 
flow that would allow the sun to produce the required luminosity more 
efficiently (i.e., in a cooler sun that derives more energy per 4p$\to ^4$He 
conversion) has a certain attractivenss.

2)\  If a flow similar to Fig. 3 were established, it is conceivable that it 
might persist, as it is both driven by and maintains the $^3$He gradient:  if 
the mixing is slowed through some perturbation, the $^3$He gradient powering 
the mixing will steepen.  The subsequent more violent ignition of $^3$He in 
the descending plumes would then act to return the cycle to equilibrium.

However this begs the question of how the cycle itself gets started.  As the 
solar spoon overstability is a large-amplitude one, the consequences for the 
static SSM are not obvious.  While a trigger for large-scale flow was discussed 
in the solar spoon [7], this has been a point of contention [8].

But a speculation made by Roxburgh [9] in a different context opens up another 
possibility:  could the difference between the SSM and a model like that 
sketched in Fig. 3 have to do with initial conditions?  The core of the early 
sun is believed to be convectively unstable prior to the establishment of 
equilibrium in the pp and CNO cycles:  $\eta$ = dlog$\epsilon$/dlogT, where 
$\epsilon$ is the energy generation rate, is initially in excess of the 
critical value of about 5.0 due to the out-of-equilibrium burning of $^{12}$C 
to $^{14}$N.  This condition persists for a time in excess of the few $\cdot 
10^7$ y characterizing the flow of Fig. 3.  The convective epoch is 
conventionally ignored in SSM simulations:  if this phase is transient, one 
can avoid a great deal of work by starting from a static, one-dimensional 
primordial sun. Roxburgh raised the issue, with 
perhaps a hint of embarrassment:  Could $^3$He transport by convective 
overshooting cause the early core mixing to persist, perhaps up to modern 
times?

This suggestion underscores a potential flaw in SSM logic.  The SSM arises 
from a one-dimensional solution of the equations describing stellar 
evolution, beginning with the assumption of a static primordial sun.  
Although the resulting modern SSM sun is affected by the solar spoon 
overstability, modelers have taken comfort in the fact that the $^3$He profile 
is only unstable to large-amplitude perturbations.  Thus it is possible $\--$ 
some might argue likely [8] $\--$ that the $^3$He overstability has no 
consequences for today's SSM sun.  Roxburgh's concern is that a more detailed 
modeling of the early convective core might allow one to evolve onto a 
trajectory where the $^3$He gradient naturally becomes involved in driving 
persistent convection.  Thus there could be a bifurcation, leading 
respectively to the SSM and to a modern convective sun that exploits the 
$^3$He gradient, associated entirely with how realistically initial conditions 
are treated.  The physics concerns raised by Refs. [7] and [9] are substantial 
ones, and I find it disturbing that our entirely phenomenological exercise in 
solar model neutrino physics leads us back to those papers.

In closing, let me stress that Andrew and I presented our work as a 
{\em highly} speculative but {\em quite} amusing possibility.  
It is not being proposed as a solution to the solar neutrino problem, but as an 
existence proof:  there do exist nonstandard model possibilities that have not 
been explored quantitatively yet could substantially reduce the problem 
summarized in Eq. (1).  For this reason I think it will remain unwise to rule 
out astrophysical solutions, at least until we have the results from Superkamiokande 
and SNO.  It is also noteworthy that only a single class of steady-state 
nonstandard models, those with core mixing on timescales characteristic of 
$^3$He mixing, appears helpful in reducing the solar neutrino discrepancies.

For those who might be motivated to pursue an astrophysical solution, the 
present work presents two challenges:

1)\  Can the ``model" be implemented dynamically?  That is, in a realistic 2D 
or 3D model with an early convective core, would the establishment of the 
$^3$He gradient allow the star to remain convective and evolve into flow patterns
 similar to Fig. 3?  This is 
quite a challenge requiring the modeling of two-fluid flows where the fluids 
are both chemically and thermally distinct.

2)\  Is such a ``model" compatible with other astrophysical constraints?  As 
mentioned in Ref. [1], core convection would likely alter galactic $^3$He 
synthesis, evolution along the color-magnitude diagram, and helioseismology.
The first might be a welcome change in galactic chemical evolution, while the 
latter two are substantial tests that a viable nonstandard model must be able 
to pass.  In fact, in a recent paper, 
John Bahcall et al. [10] have argued that such core mixing models are 
definitively ruled out by helioseismology.  I agree with John that the SSM 
helioseismology successes are very significant.  But I regard the conclusions 
about mixed models to be premature.  Here and in Ref. [1] it is clear that we 
have not provided a model, but merely sketched a somewhat provocative idea.  
Thus the use of quotation marks on ``model".  The work of Ref. [10] rules out 
a ``model" analogous to the static cartoon of Fig. 2.  Such a model is also 
ruled out by $\vec F = m (r) \vec a$, as no attempt has been made to enforce the 
standard rules of stellar hydrodynamics.  Likewise, were the SSM described in 
a similarly crude fashion, it would almost certainly fail an analogous
helioseismology test.  My point is that if we succeed in item 1) above - if a 
detailed model can be constructed - we will then have the density and 
temperature profiles necessary for determining sound speeds c(r) and 
helioseismology.  Helioseismology will then be both an appropriate test of the 
model, and a major challenge to its viability. I would purpose such a 
statement as a more conservative summary of the work reported in Ref. [10].
\section*{Acknowledgments}
This work was supported in part by the U.S. Department of Energy.  The efforts 
of Andrew Cumming, my collaborator in Ref. [1], were supported by University 
of Washington Research Experiences for Undergraduates program, funded by the 
National Science Foundation and the Institute for Nuclear Theory, 
University of Washington.

\end{document}